\documentclass{webofc}
\usepackage[varg]{txfonts}   
\usepackage{lineno}
\newcommand{\pT}{\ensuremath{p_{\rm T}}\xspace}
\newcommand{\pt}{\pT}

\newcommand{\TeV}{\ensuremath{{\rm TeV}}\xspace}

\newcommand{\ppi}{\ensuremath{\pi}\xspace}
\newcommand{\ppipm}{\ensuremath{\ppi^{\pm}}\xspace}
\newcommand{\pka}{\ensuremath{K}\xspace}
\newcommand{\pkapm}{\ensuremath{\pka^{\pm}}\xspace}
\newcommand{\pphi}{\ensuremath{\phi}\xspace}
\newcommand{\ppr}{\ensuremath{p}\xspace}
\newcommand{\pprpm}{\ensuremath{\ppr, \bar{\ppr}}\xspace}
\newcommand{\pbpb}{\ensuremath{\text{Pb--Pb}}\xspace}
\newcommand{\xexe}{\ensuremath{\text{Xe--Xe}}\xspace}

\newcommand{\snn}{\ensuremath{\sqrt{s_{\rm NN}}}\xspace}

\newcommand{\enTeV}[1]{\ensuremath{#1~\TeV}\xspace}

\newcommand{\snncTeV}{\ensuremath{\snn~=~\enTeV{5.02}}\xspace}
\newcommand{\snncqTeV}{\ensuremath{\snn~=~\enTeV{5.44}}\xspace}
\newcommand{\snndTeV}{\ensuremath{\snn~=~\enTeV{2.76}}\xspace}

\newcommand{\vtwo}{\ensuremath{v_{2}}\xspace}
\newcommand{\vtwotwo}{\ensuremath{\vtwo\{2, |\Delta\eta|>2\}}\xspace}
\newcommand{\avdndeta}{\ensuremath{\langle{\rm d}N_{\rm ch}/{\rm d}\eta\rangle}\xspace}
\begin{document}
\title{Studying light flavour hadrons produced in the collision of different nuclei at the LHC}
%
%

\author{\firstname{Nicol\`o} \lastname{Jacazio}\inst{1}\fnsep\thanks{\email{nicolo.jacazio@cern.ch}} For the ALICE Collaboration.
}

\institute{CERN
}

\abstract{%
The study of identified particle production as a function of event multiplicity is a key tool for understanding the similarities and differences among different colliding systems.
Now for the first time, we can investigate how particle production is affected by the collision geometry in heavy-ion collisions at the LHC.
In these proceedings, we report newly obtained ALICE results on charged and identified particle production in \pbpb and \xexe collision at $\snn = 5.02$ and $\snn = 5.44$ TeV, respectively, as a function of transverse momentum (\pt) and collision centrality.
Particle spectra and ratios are compared between two different colliding systems at similar charged-particle multiplicity densities (\avdndeta), and different initial eccentricities.
We find that in central collisions, spectral shapes of different particles are driven by their masses.
The \pt-integrated particle yield ratios follow the same trends with \avdndeta as previously observed in other systems, further suggesting that at the LHC energies, event hadrochemistry is dominantly driven by the charged-particle multiplicity density and not the collision system, geometry or center-of-mass energy.}
\maketitle
\section{Introduction}
\label{intro}
The ultimate goal of heavy-ion physics is the study of the properties of the Quark-Gluon Plasma (QGP), a deconfined and chirally restored state of matter.
The measurement of the transverse momentum (\pT) spectra of identified particles provides a way into the collective properties and of particle production in the fireball created in heavy-ion collisions.
The ALICE experiment \cite{Aamodt:2008zz, Abelev:2014ffa} is particularly well suited to study the production of both identified and unidentified charged particles due to its excellent tracking performance coupled with extensive particle identification (PID) capabilities over a wide range of transverse momentum.
Particles are identified by combining different techniques allowing their continuous separation over a large transverse momentum interval.
In 2017 ALICE recorded \xexe collision at the highest energy per nucleon ever achieved in the laboratory, allowing for the first time at the LHC a quantitative comparison of different heavy-ion collision systems.

\section{Results}
\label{sec-1}
We report results on the production of \ppi, \pka, \ppr and \pphi measured in \xexe collisions at \snncqTeV as a function of centrality, as discussed in~\cite{ALICE:2021lsv}.
The data sample was recorded with a minimum-bias trigger.
The total charge collected in the V0 detectors, a set of two scintillator detectors located in the pseudorapidity region $2.8 < \eta < 5.1$ (V0A) and $- 3.7 < \eta < - 1.7$ (V0C) and covering the full azimuth, was used to determine the centrality of each \xexe collision defined as percentiles of the total hadronic cross section.
Further details are given in \cite{ALICE-PUBLIC-2018-003}.
Contributions from weak decays of strange particles and from particle knock-out in the material were removed with the data driven approach described in \cite{ALICE:2021lsv}.
Systematic uncertainties were estimated by varying the PID technique and the selection criteria used to define the track sample.
The amount of pile-up per event was reduced by selecting runs with low interaction rate and by rejecting events with more than one reconstructed vertex, resulting in a negligible effect.
The evaluation of the efficiency and acceptance corrections was performed using events simulated with the HIJING \cite{HIJING} event generator and embedded into a detailed description of the ALICE detector through which tracks are propagated with the GEANT3 \cite{GEANT3} transport code.
\begin{figure}
  \centering
  \includegraphics[width=.7\textwidth]{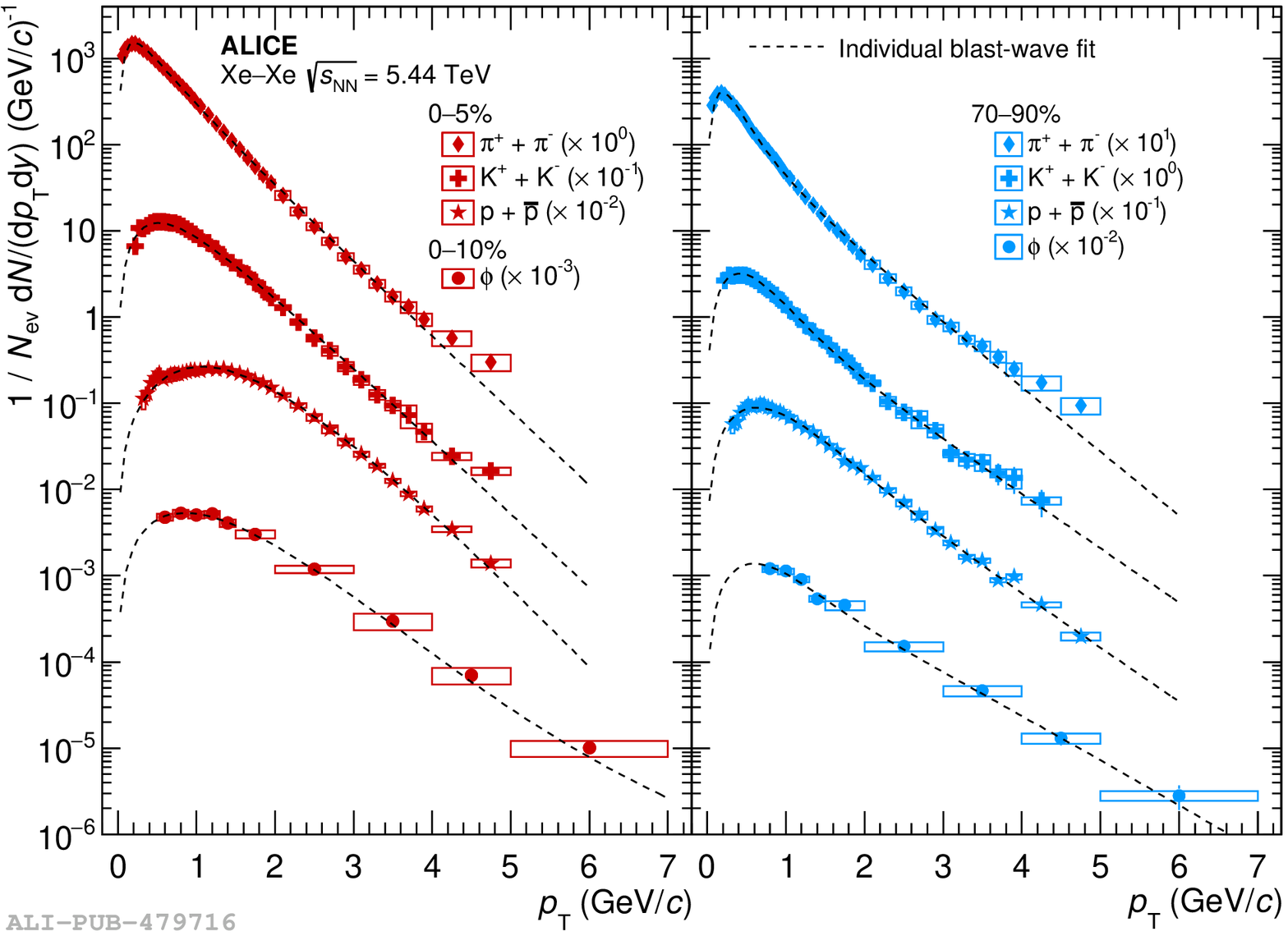}
  \caption{
    Transverse momentum distributions of \ppipm, \pkapm, \pprpm and \pphi as measured in central (left) and peripheral (right) \xexe collisions at \snncqTeV.
  }
  \label{figA}
\end{figure}
\begin{figure}
  \centering
  \begin{minipage}[c]{0.53\textwidth}
    \includegraphics[width=\textwidth]{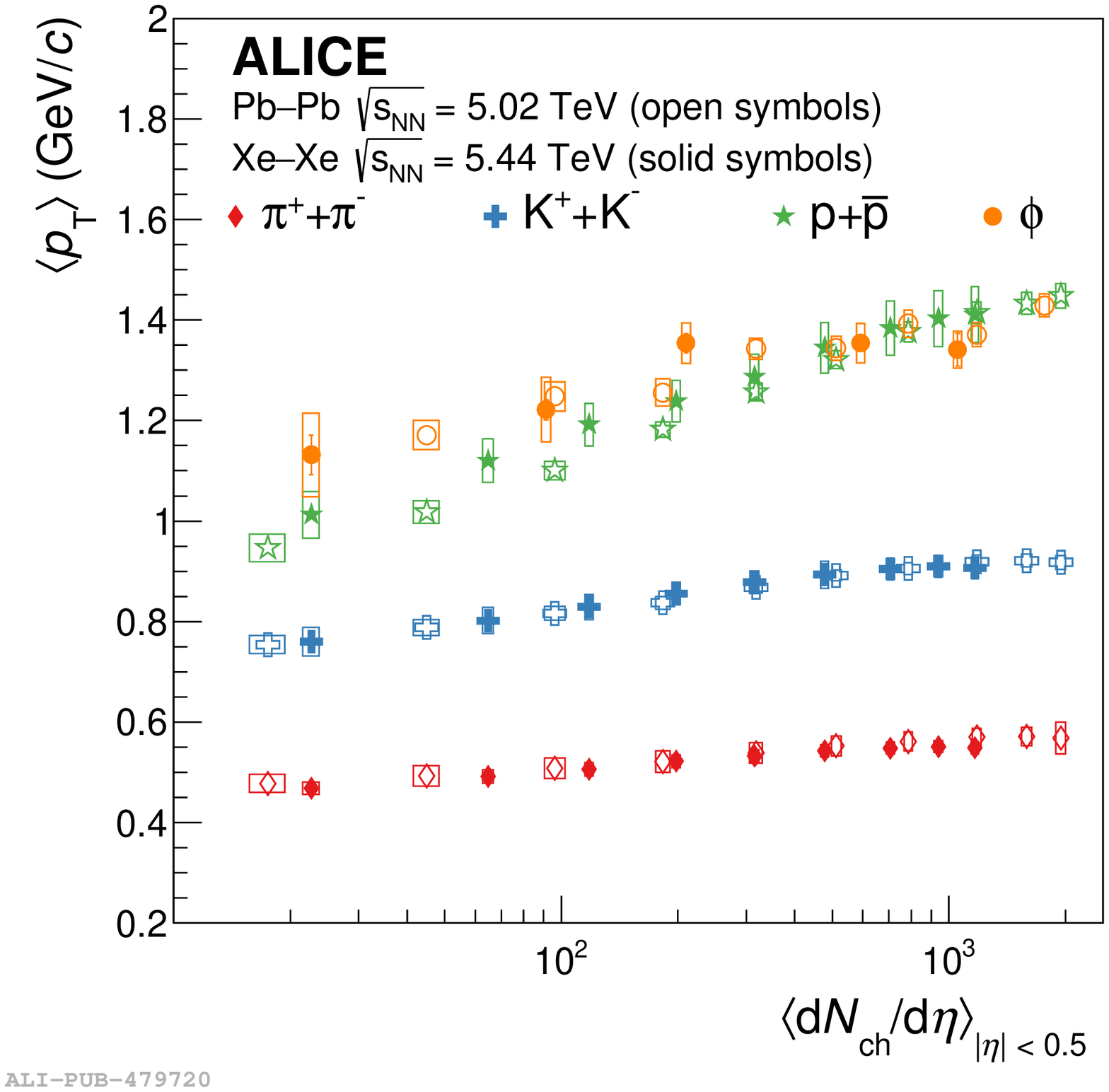}
  \end{minipage}\hfill
  \begin{minipage}[c]{0.39\textwidth}
    \caption{
      Average transverse momentum of \ppipm, \pkapm, \pprpm and \pphi as a function of the charged-particle multiplicity density as measured in \xexe collisions at \snncqTeV \cite{ALICE:2021lsv} and compared to \pbpb results at \snncTeV \cite{Acharya:2019qge, PhysRevC.101.044907}.
      }
      \label{figB}
    \end{minipage}
\end{figure}
\begin{figure}
  \centering
  \begin{minipage}[c]{0.72\textwidth}
  \includegraphics[width=\textwidth]{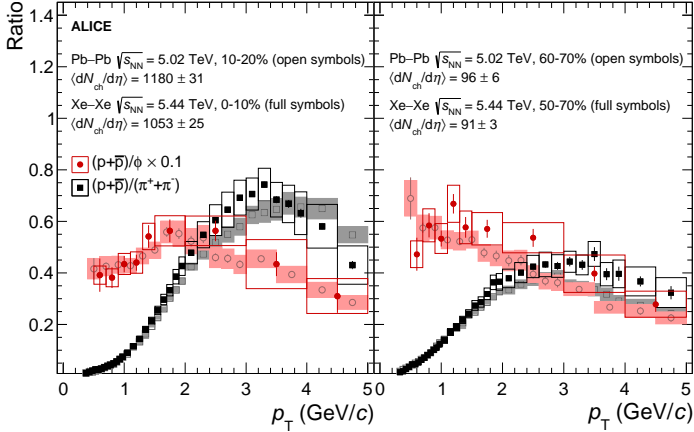}
\end{minipage}\hfill
\begin{minipage}[c]{0.27\textwidth}
  \caption{
    Left: $\ppr/\pphi$ and $\ppr/\ppi$ as a function of the transverse momentum in 0--10\% central \xexe collisions at \snncqTeV \cite{ALICE:2021lsv} compared to 10--20\% central \pbpb collisions at \snncTeV \cite{Acharya:2019qge, PhysRevC.101.044907} to match the charged particle multiplicity in the two systems.
    The same is shown in the right panel for peripheral collisions.
    }
    \label{figC}
  \end{minipage}
\end{figure}
\begin{figure}
  \centering
  \begin{minipage}[c]{.45\textwidth}
    \includegraphics[width=\textwidth]{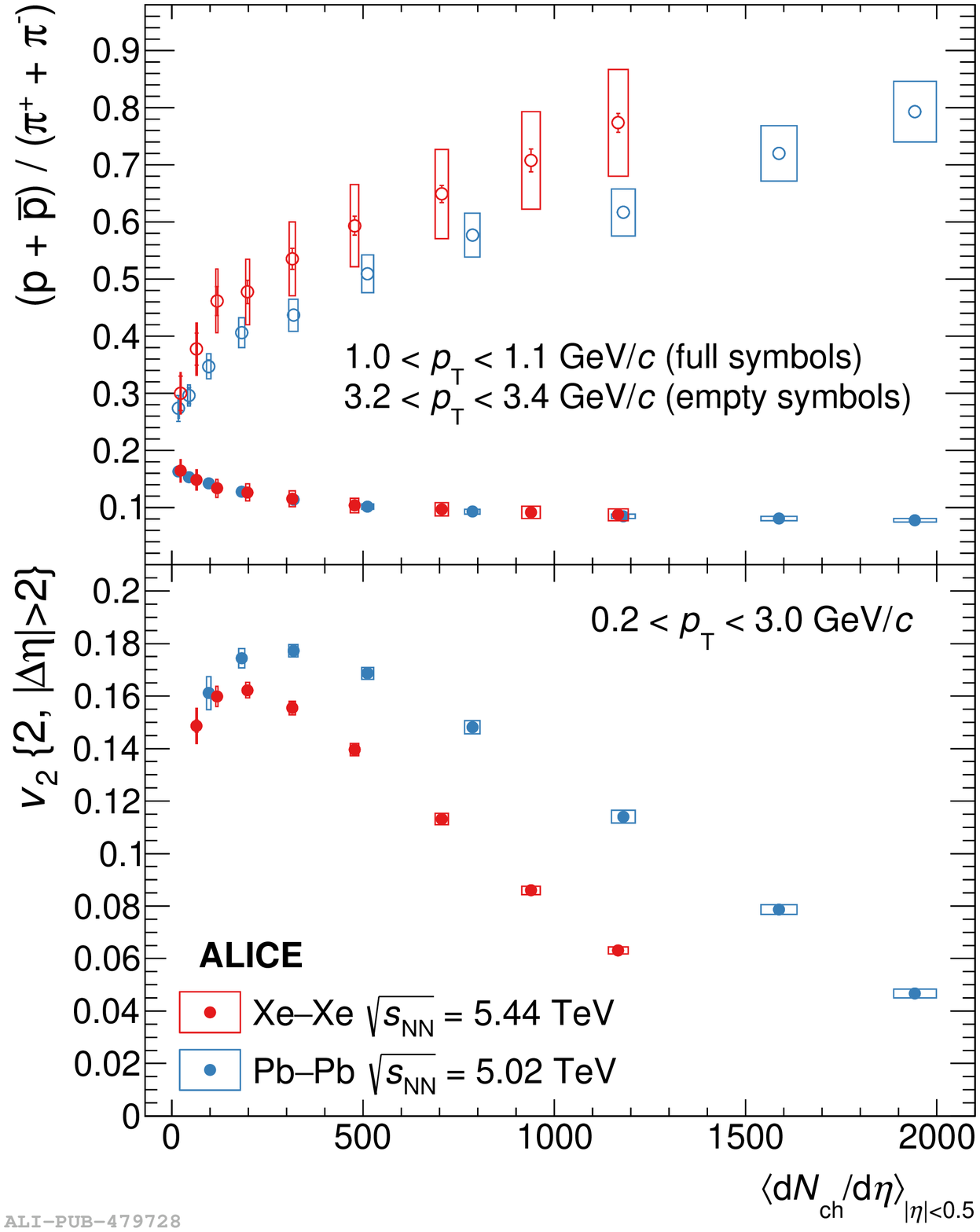}
  \end{minipage}\hfill
  \begin{minipage}[c]{.45\textwidth}
    \includegraphics[width=\textwidth]{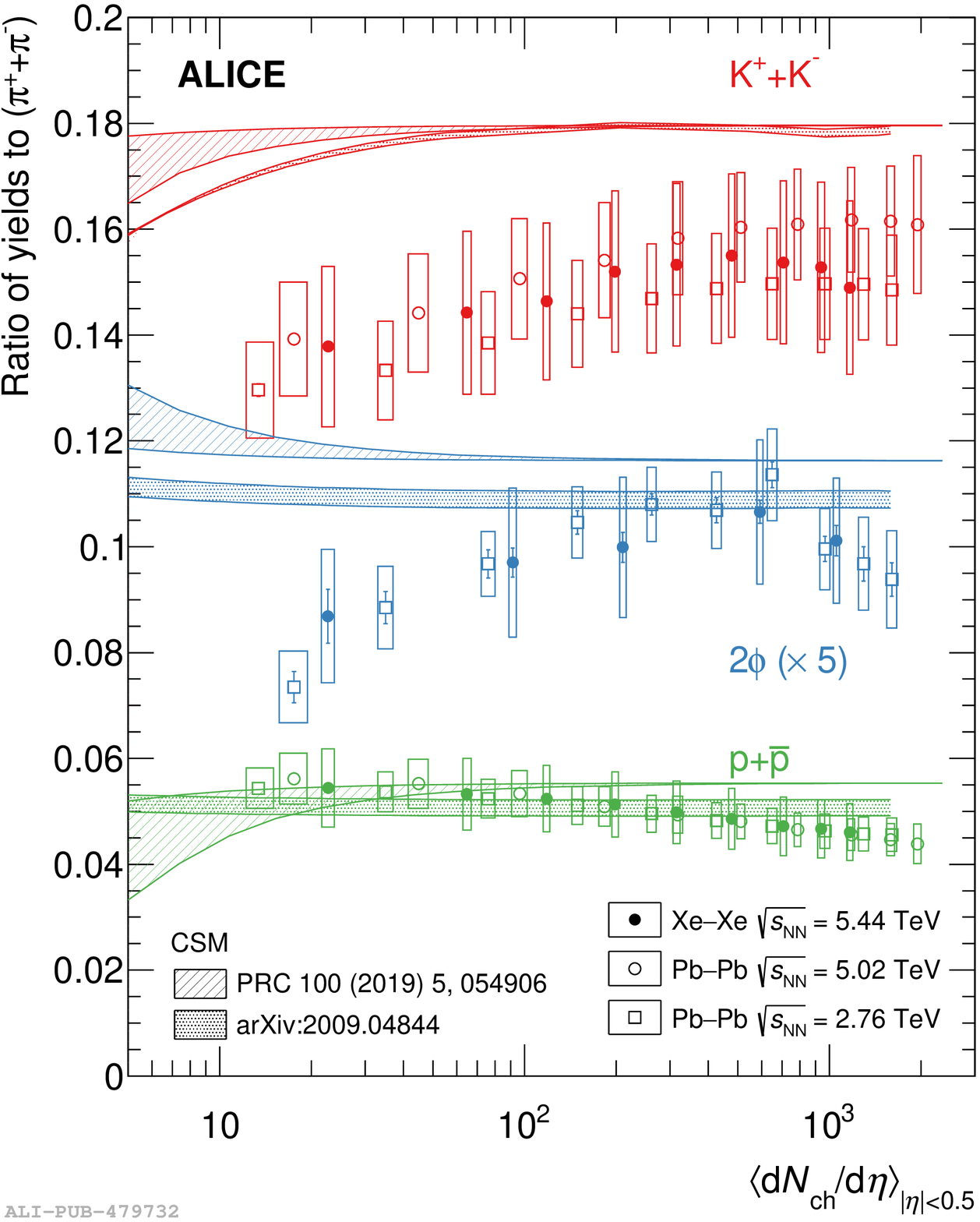}
  \end{minipage}
  \caption{
    Left figure, top panel: $\ppr/\ppi$ as a function of charged particle multiplicity density in two \pt intervals for \xexe \snncqTeV \cite{ALICE:2021lsv} compared to \pbpb collisions at \snncTeV~\cite{PhysRevC.101.044907, Acharya:2019qge}.
    Left figure, bottom panel: the flow coefficient \vtwotwo reported for the same collision systems~\cite{Acharya:2018zuq, Acharya:2018ihu} as a function of charged particle multiplicity density.
    Right figure: $\pka/\ppi$, $\ppr/\ppi$ and $\ppr/\ppi$ for integrated yields as a function of the charged-particle multiplicity density for \xexe collisions at \snncqTeV \cite{ALICE:2021lsv} compared to \pbpb collisions at \snndTeV~\cite{Abelev:2013vea, Abelev:2014uua} and 5.02~TeV~\cite{PhysRevC.101.044907, Acharya:2019qge}.
    Predictions from the canonical statistical model (CSM) are shown as bands considering different correlation volumes~\cite{Vovchenko:privatecomm} (based on \cite{Vovchenko:2019kes}) and chemical freeze-out temperatures~\cite{Cleymans:2020fsc}.
  }
  \label{figD}
\end{figure}
The \pT spectra of identified \ppi, \pka, \ppr and \pphi are shown for different centrality classes in Fig.~\ref{figA}.
A visual comparison of the spectral shapes reveals that the spectra become harder with increasing centrality.
This hardening is found to be mass dependent with protons being more affected than pions and experiencing a similar effect as the \pphi meson.
This is consistent with the presence of a collective evolution, known also as radial flow.
This effect can be better appreciated in Fig.~\ref{figB} where the average transverse momentum extracted from the spectra is shown as a function of the charged particle multiplicity for each particle and compared to the results obtained in \pbpb collisions.
The increase in average \pt is the largest for the heavier particles.
In addition, it is striking to observe that in the two systems the average \pt are similar provided that they are measured at the same multiplicity and particles with similar mass i.e. \ppr and \pphi have similar average \pt.
The effect of radial flow can also be seen in the baryon/meson ratios shown in Fig.~\ref{figC} as a depletion of $\ppr/\ppi$ at low \pt and an enhancement at higher \pt.
It is worth noting that the baryon/meson ratio exhibits an almost flat behavior only when considering particles with similar mass (i.e. \ppr and \pphi) indicating that the hardening is driven by the mass and not by the quark content of the hadron.
The $\ppr/\ppi$ ratio is shown as a function of multiplicity in Fig.~\ref{figD} (left) to further illustrate the effect of the radial flow, here it can be seen that the amount of enhancement (depletion) at high \pt (low \pt) is compatible in \xexe and \pbpb collisions when compared at similar multiplicity.
Conversely, this is not true for the \vtwo that exhibits a strong dependence on the centrality, indicating a dominant effect of the initial shape of the collision.
Also hadrochemistry is found to be scaling well with the charged particle multiplicity as it can be seen in Fig.~\ref{figD} (right).
Here a general agreement of the results is found across different systems and energies.
Furthermore, the results are found to be in agreement with predictions from thermal models.

\section{Conclusions}
\label{sec-2}
In these proceedings, results on the \ppipm, \pkapm, \pprpm and \pphi production measured as a function of centrality in \xexe collisions at \snncqTeV were reported~\cite{ALICE:2021lsv}.
For the first time at the LHC, it was possible to compare collisions of different nuclei.
A comparison of data across different collision systems indicates that the features of light hadron production scale with the event multiplicity and are independent of the colliding system, collision geometry or collision energy.
In addition, the radial flow is correlated with the charged particle multiplicity, and the anisotropic flow is affected by collision geometry.
It was also shown that the hadrochemistry does not depend significantly on the collision system and scales with the charged particle multiplicity.
%
\bibliography{bibliography.bib}
%
%

\end{document}